# High critical current density and low anisotropy in textured $Sr_{1-x}K_xFe_2As_2$ tapes for high field applications

Zhaoshun Gao, Yanwei Ma*, Chao Yao, Xianping Zhang, Chunlei Wang, Dongliang Wang

*Key Laboratory of Applied Superconductivity, Institute of Electrical Engineering, Chinese Academy of Sciences, Beijing 100190, China*

Satoshi Awaji, Kazuo Watanabe

*High Field Laboratory for Superconducting Materials, Institute for Materials Research, Tohoku University, Sendai 980-8577, Japan*

From the application point of view, large critical current densities $J_c$ (H) for superconducting wires are required, preferably for magnetic fields higher than 5 T. Here we show that strong $c$-axis textured $Sr_{1-x}K_xFe_2As_2$ tapes with nearly isotropic transport $J_c$ were fabricated by an ex-situ powder-in-tube (PIT) process. At 4.2 K, the $J_c$ values show extremely weak magnetic field dependence and reach high values of $1.7 \times 10^4$ A/cm$^2$ at 10 T and $1.4 \times 10^4$ A/cm$^2$ at 14 T, respectively, these values are by far the highest ever reported for iron based wires and approach the $J_c$ level desired for practical applications. Transmission electron microscopy investigations revealed that amorphous oxide layers at grain boundaries were significantly reduced by Sn addition which resulted in greatly improved intergranular connectivity. Our results demonstrated the strong potential of using iron based superconductors for high field applications.

---

* Author to whom correspondence should be addressed; E-mail: ywma@mail.iee.ac.cn

The discovery of superconductivity in the iron pnictides has generated a great deal of research interest, not only in basic physics, but also in the field of applied superconductivity [1-7]. In addition to their high transition temperature, $T_c$, the Fe based superconductors were reported to have a rather high upper critical field, $H_{c2}$, and low $H_{c2}$ anisotropy [8–11]. Data reported so far has revealed many similarities between iron pnictide and cuprate superconductors, such as a layered crystalline structure and superconductivity induced by carrier doping. However, there are some distinct differences, such as a highly symmetric order parameter based on $s+$- wave in the iron pnictides compared to a $d$-wave pairing [12]. Therefore, different characteristics of high-angle GBs can be expected. Indeed, Katase et al. [13] performed a systematic study on the misorientation dependence of inter-grain $J_c$ using an epitaxially grown Ba122 bi-crystal. The critical angle ($\theta_c$) of the transition from a strong link to a weak link for Ba122 was found to be 9°, which is substantially larger than the value of 3~5° reported for YBCO. A remarkable transport inter-grain $J_c$ of ~$10^5$ A/cm$^2$ at 4 K and self field has been observed in samples of high misorientation angle of 45°. Such excellent properties suggest that the Fe based superconductors are promising for use in high magnetic field generation at liquid helium.

Earlier results indicate that the global critical current is limited by intergrain currents across grain boundaries in iron based superconductors [13–21]. Some of which can be ascribed to extrinsic factors such as porosity, the amorphous phase at grain boundaries, and micro-cracks. Concerning the main intrinsic factor, $J_c$ across the grain boundary decreases with grain boundary misorientation angle similar to that observed in the cuprates YBCO [13, 21]. In order to improve the grain connectivity of the iron based superconductors, a number of experimental techniques, including metal additions [22-25], rolling induced texture process [26, 28], and hot isostatic press method [27], have been attempted. In this work, we report the further improvement of the transport critical current properties in textured $Sr_{0.6}K_{0.4}Fe_2As_2$ tapes with Sn additions. We observed a high transport critical current density of >10 kA/cm$^2$ at 4.2 K in 14 T field. Such a value is by far the highest ever reported for Fe based wires. The influences of Sn addition, microstructure, and the annealing process on achieving superior $J_c$ properties will be discussed.

**Results**

The degree of grain texture was determined by XRD analysis. X-ray diffraction patterns of $Sr_{1-x}K_xFe_2As_2$ bulk precursor, as-rolled tape, low temperature annealed tape, and high temperature annealed tape are shown collectively in Figure 1. The relative intensity of (103) peak with respect to that of (002) peak in all tape samples, when compared to the bulk precursor, is strongly reduced, indicating enhanced $c$-axis orientation. In order to have quantitative information about the texture of the $Sr_{1-x}K_xFe_2As_2$ phase, we have evaluated the $c$ axis orientation factor F by the Lotgering method as follows [29].

$$F=(\rho - \rho_0)/(1 - \rho_0),$$

Where $\rho = \sum I(00l)/ \sum I(hkl)$, $\rho_0 = \sum I_0(00l)/ \sum I_0(hkl)$, $I$ and $I_0$ are the intensities of each reflection peak ($hkl$) for the oriented and random samples, respectively. The value of F for the as-rolled tape, was 0.316, the value of F for low temperature annealed tape was 0.348, and the value of F for high temperature annealed tape was 0.565, respectively. The cold deformation of the $Sr_{1-x}K_xFe_2As_2$ tapes during the tape fabrication processes induces a relevant texture. In particular, the texture greatly improved after high temperature heat treatments.

Figure 2 presents the temperature dependent magnetic moment of samples measured in the magnetic field parallel to the $c$-axis. It is evident from the zero-field cooled (ZFC) signal that the diamagnetic signal does not saturate but continuous to decrease to lower temperature indicating some degrees of inhomogeneity in the sample annealed at high temperature. Note that low temperature annealing with longer time sharpens the $T_c$ transition and reduces the inhomogeneities. Additionally, as shown more clearly in figure 2, low temperature annealing increased the onset $T_c$ from 27.3 K to 32.5 K.

High critical current density in applied magnetic fields is essential for wire applications. Figure 3 shows the transport $J_c$ measured as a function of applied magnetic fields along with reported values of other Fe-based superconducting wires and conventional Nb based superconducting wires [25, 27]. For the Batch II samples heated at low temperature, only data above 2 T are shown, because at lower field region, $I_c$ is too high to be measured with our current apparatus. It can be seen that both samples demonstrate excellent $J_c$ performance. Remarkably, the $J_c$ of the tapes sintered at low temperature exhibits very weak field dependence and maintains a reasonably high value of $1.4 \times 10^4$ A/cm$^2$ at 14 T, this $J_c$ value is the highest ever reported for the iron based superconducting wires or tapes so far. It is about a factor of two higher than the best Ba-122 wire recently reported in ref. [27]. Another interesting feature is the comparison of the superconducting properties of the present tapes with those of classical superconductors. The $J_c$ value of $1.7 \times 10^4$ A/cm$^2$ was achieved at 10 T for the tapes sintered at low temperature, which exceeded the value of NbTi conductors [25]. To compare with the $J_c$ of Nb$_3$Sn, the data above 14T for Batch II in Fig.3 is extrapolated from low fields. As can be seen, a crossover with the $J_c$ of Nb$_3$Sn would take place at around 19.5 T, as shown in figure 3. This indicates that the 122 superconducting wires may be competitive with well established Nb based conductors used in MRI and NMR for high field generation.

For application in superconducting magnets, low anisotropy in critical current ($I_c$) is more desirable. Figure 4 shows the $I_c(H//ab)$ and $I_c(H//c)$ dependences at 4.2 K for typical Batch I and II samples. For both cases, $I_c$ is larger when the applied field is perpendicular to the tape surface instead of along the *ab*-plane. Similar results were reported in several textured 122 films [30-32]. In particular, in both of our samples, the anisotropy ratio of $I_c$ ($\gamma=I_{c\perp}/I_{c//}$ <2) appears to be much less than that in the cuprate superconductors. More interestingly, the $I_c$ in Batch II tapes is nearly isotropic, which is a highly desirable property for high field applications.

Low magnification TEM images showed a grain boundary network in the textured Sr$_{0.6}$K$_{0.4}$Fe$_2$As$_2$ tape at, which are not shown here. The tape sample exhibits a layered structure with good alignment. Figure 5 (a) shows EDX line scan across one of grain boundary. Previous studies reported [15, 16, 33] that the grain boundaries are usually coated by amorphous oxide layers. In contrast, we observed that the grain boundaries in Sn doped tapes are often filled with Sn rich materials (see fig.5 (b)), note that the thickness of these interfaces are around 2-3 nm. Clearly, Sn and its alloy as flux can improve the crystallization of grain boundaries, diminish the formation of amorphous layer, and hence improve intergranular connectivity. In order to investigate the effect of annealing process on the tapes, we studied the difference in microstructures of the tapes heated at different temperatures. Figure 5 (c) and (d) present scanning microscopy photomicrographs of Batch I and II tapes. Well-developed texture structures can be seen in both samples.

**Discussion**

As mentioned above, the $J_c$ values of textured 122 pnictide tapes have reached over $10^4$ A/cm$^2$ in a field of 14 T at 4.2 K. The well developed grain texture in our samples is a main reason for the superior $J_c$ performance, since the misalignment in crystalline orientation at grain boundaries is a crucial weak-link in connectivity. Secondly, as shown by TEM study, Sn can promote the crystallization at grain boundaries, diminish the formation of amorphous layer, and hence much improved intergranular connectivity. Thirdly, low temperature synthesis results in highly dense material, which further contributes to good connectivity.

Our present results clearly show that textured Sr122 superconducting tapes, fabricated by the low-cost powder-in-tube process, exhibited extremely high and nearly isotropic transport $J_c$, as well as independent of the field behavior, demonstrating the potential for high field magnet applications. As the cumulative knowledge of pnictide PIT process grows, it is expected that the critical currents will continue to increase. We believe that the current PIT process can be applied industrially to fabricate high performance pnictide wires and tapes, as already demonstrated by the production of Bi-2223 and MgB$_2$ wires and tapes.

In summary, excellent transport $J_c$ values under high magnetic field were observed in the textured Sr$_{1-x}$K$_x$Fe$_2$As$_2$ superconducting tapes that exhibit very small anisotropy. The well aligned superconducting grains and strengthened intergrain coupling achieved by Sn addition are responsible for this high $J_c$ performance. With further improvement of critical current density and wire fabrication technology, use of $Sr_{1-x}K_xFe_2As_2$ for very high field application is feasible.

## Methods

**Sample preparation.** The $Sr_{1-x}K_xFe_2As_2$ polycrystalline precursors were prepared by a one-step PIT method developed by our group [28]. In order to compensate for loss of elements during the milling and sintering procedures, the starting mixture contains 10-20% excess K. The precursors were ground to a powder in an agate mortar and pestle. For Sn doped samples, 5-10 wt% Sn was added to the precursor powder and then the mixture was ground in a mortar for half an hour. The final powders were filled and sealed into an iron tube, which was subsequently swaged and drawn down to a wire of 1.9 mm in diameter. The as-drawn wires were then cold rolled into tapes (0.6 mm in thickness) with a reduction rate of 10~20%. The tapes were finally sintered following two different processes: sample was directly inserted into a furnace that was preheated to 1100°C, and removed from the furnace after 1~30 minutes (high temperature annealing process: Batch I), sample was sintered at 800-950°C for 1~30 minutes and then decreased to 600°C for 5 h (low temperature annealing process: Batch II).

**Measurements.** Phase integrity and texture of $Sr_{0.6}K_{0.4}Fe_2As_2$ grains were investigated using x-ray diffraction (XRD), for which iron sheath was mechanically removed after cutting the edges of the tape. Microstructure was studied using a scanning electron microscopy (SEM). DC magnetization measurements were carried out using a physical property measurement system (PPMS). The transport current $I_c$ at 4.2 K and its magnetic field dependence were evaluated at the High Field Laboratory for Superconducting Materials (HFLSM) at Sendai, by standard four-probe method, with a criterion of 1 μV/cm, then the critical current was divided by the cross section area of the superconducting core to get the critical current density $J_c$. For each set of tapes, $I_c$ measurement was performed on several samples to ensure the reproducibility.

## Acknowledgements

The authors thank Drs. Qingxiao Wang and XiXiang Zhang at the KAUST for the TEM investigation. They are also indebted to Dr. Ma at ANL for useful suggestion. This work is partially supported by the National '973' Program (Grant No. 2011CBA00105) and National Natural Science Foundation of China (Grant No. 51002150 and 51025726).

## Author contributions

Z.S.G. and C. Y. did most of the synthesis and characterizations. X.P.Z. and C. Y. did the heat treatment and in-field $I_c$ measurement experiments. Z.S.G. and Y.W. M. wrote the paper. C. L.W., D. L. W., S.A. and K.W. helped with the experiment.

## Additional information

Competing financial interests: The authors declare no competing financial interests.

# Captions

Figure 1  X-ray diffraction patterns for $Sr_{1-x}K_xFe_2As_2$ precursor bulk (a), as-rolled tape (b), tape annealed at low temperature (c) and tape annealed at high temperature (d).

Figure 2  Temperature dependence of magnetic susceptibility measured with field of 20 Oe for Batch I and Batch II samples.

Figure 3  The transport $J_c$ values at 4.2 K obtained in this experiment plotted as a function of applied magnetic fields along with other Fe-based superconducting wires. The conventional Nb based and Bi2212 superconducting wires are also included for reference.

Figure 4  Magnetic field dependence of $I_c$ at 4.2 K for $Sr_{0.6}K_{0.4}Fe_2As_2$ tapes annealed at different temperatures. The magnetic field was applied parallel and perpendicular to the tape surface.

Figure 5  (a) A STEM image of the $Sr_{0.6}K_{0.4}Fe_2As_2$ grain boundary. (b) The EDS line scan (as denoted in (a) ) indicates that the grain boundary is filled with Sn rich material. Typical SEM images for $Sr_{0.6}K_{0.4}Fe_2As_2$ tapes annealed at high temperature (c) and low temperature (d).

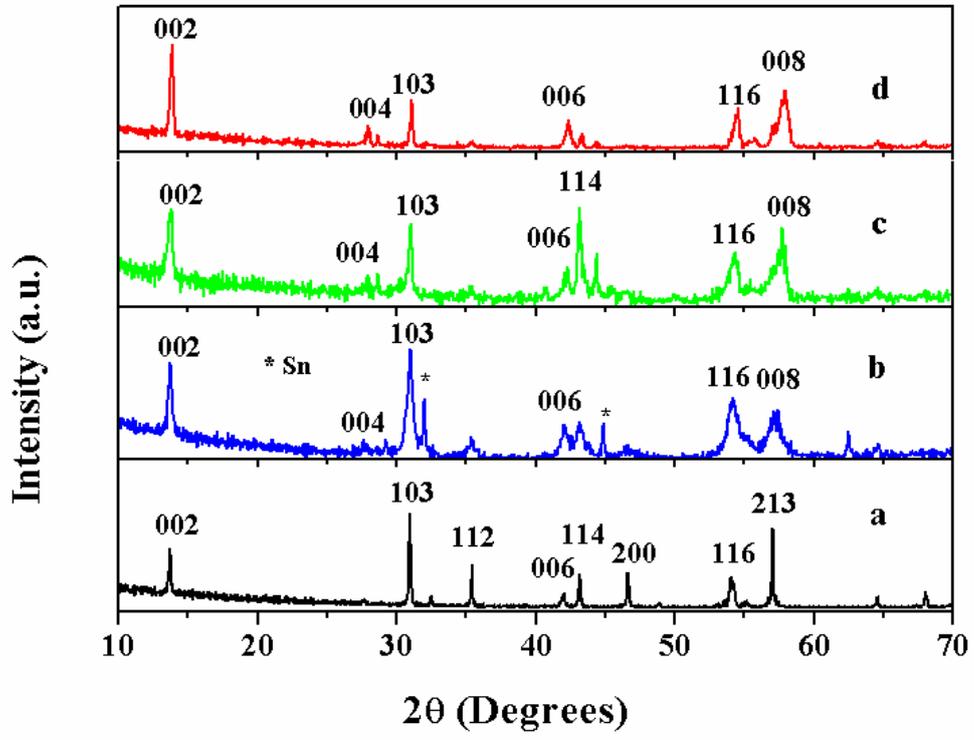

Figure 1

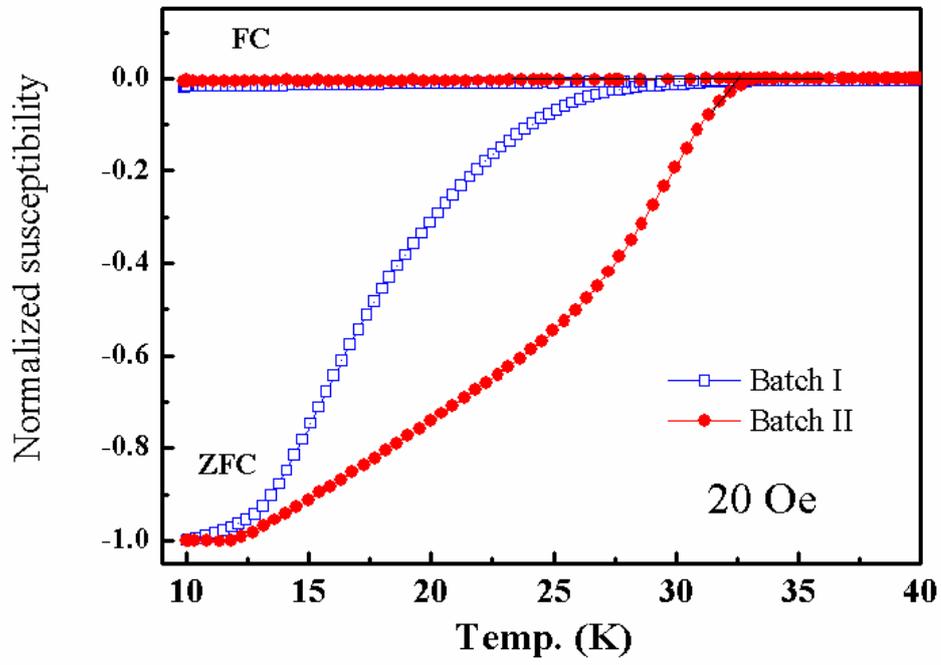

Figure 2

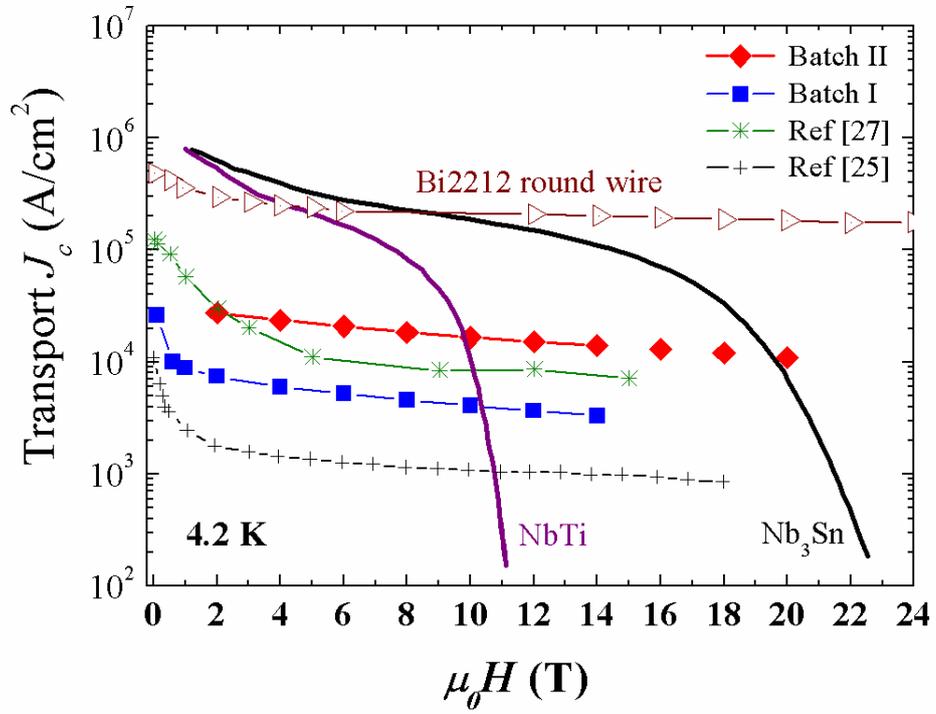

Figure 3

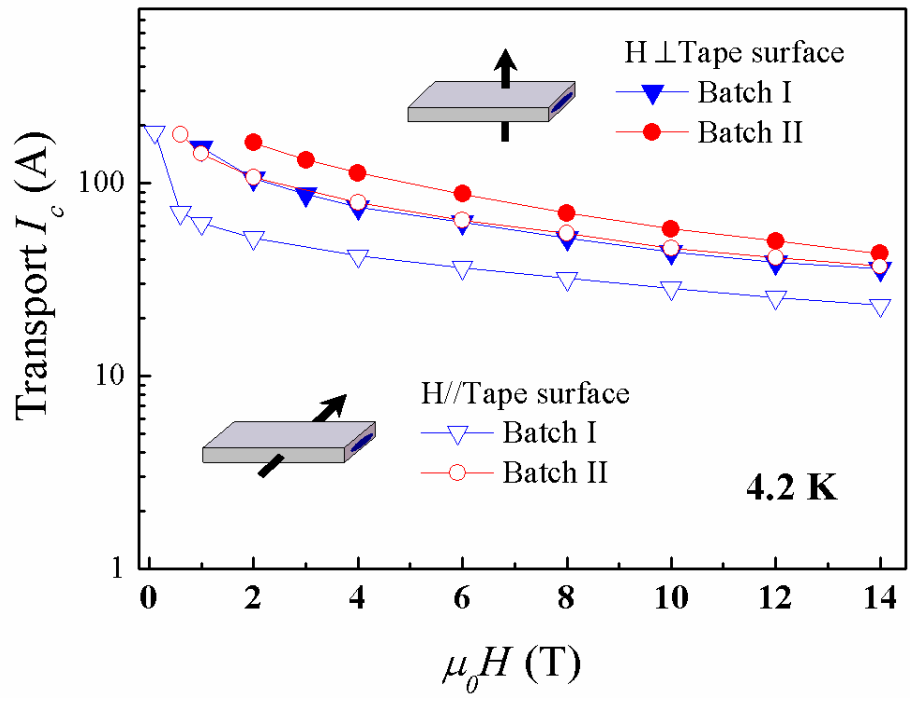

Figure 4

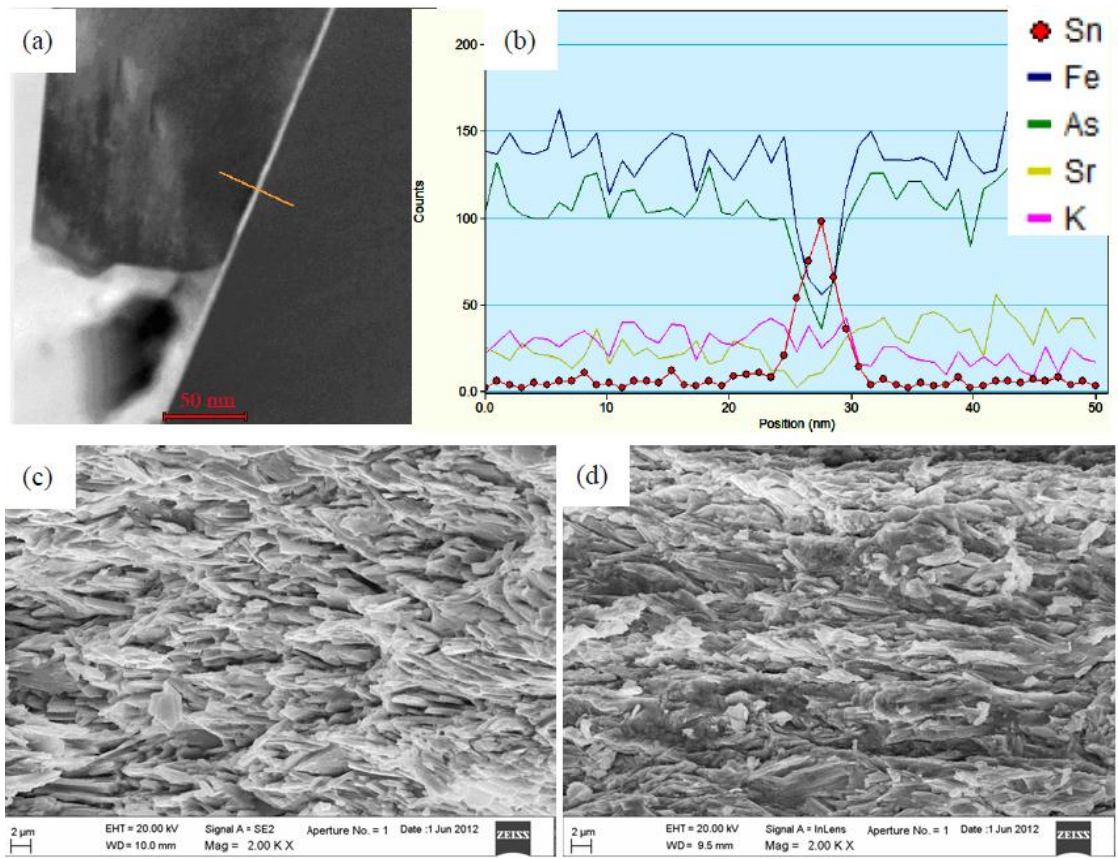

Figure 5